\documentclass[a4paper,11pt]{article}
\pdfoutput=1
\usepackage{jcappub}
\usepackage{tikz,xcolor,hyperref}
\usepackage{amsmath, amssymb, amsthm, graphicx, epsfig, fancyhdr,epsfig, slashed}
\usepackage[normalem]{ulem}
\usepackage{tikzsymbols}
\usepackage{natbib}
\usepackage{float}
\usepackage{placeins}

\def\beq{\begin{equation}}
\def\eeq{\end{equation}}
\def\bea{\begin{eqnarray}}
\def\eea{\end{eqnarray}}
\def\be{\begin{equation}}
\def\ee{\end{equation}}

\def\bse{\begin{subequations}}
\def\ese{\end{subequations}}

\def\Mpl{M_{P}}
\def\f{\frac}
\def\l{\left}
\def\r{\right}

\def\Tbh{T_{\text {BH}}}
\def\Tev{T_{\text {ev}}}

\def\tin{t_{\rm in}}

\def\Min{M_{\text {in}}}

\def\mbh{M_{\text {BH}}}

\def\mdm{M_{\text {DM}}}

\def\tin{t_{\rm in}}

\def\Tin{T_\text{in}}
\def\mdm{m_\text{DM}}
\def\gss{g_{\star s}}
\def\gs{g_{\star}}

\def\tev{t_{\rm ev}}
\def\Qin{Q_{\rm in}}
\def\ldm{\lambda_{\rm DM}}
\def\Ldm{\Lambda_{\rm DM}}
\def\qdm{q_{\rm DM}}
\def\Qdm{Q_{\rm DM}}
\begin{document}
\title{Asymmetries from a  charged memory-burdened PBH}
\author[a]{Basabendu Barman,}
\emailAdd{basabendu.b@srmap.edu.in}
\author[b]{Kousik Loho}
\emailAdd{kousikloho@hri.res.in}
\author[c]{and Óscar Zapata}
\emailAdd{oalberto.zapata@udea.edu.co}
\affiliation[a]{\,\,Department of Physics, School of Engineering and Sciences, SRM University-AP, Amaravati 522240, India}
\affiliation[b]{\,\,Regional Centre for Accelerator-based Particle Physics, Harish-Chandra Research Institute, A CI of Homi Bhabha National Institute, Chhatnag Road, Jhunsi, Prayagraj 211019, India}
\affiliation[c]{\,\,Instituto de Física, Universidad de Antioquia\\Calle 70 \# 52-21, Apartado Aéreo 1226, Medellín, Colombia}
\abstract{We explore a purely gravitational origin of observed baryon asymmetry and dark matter (DM) abundance from asymmetric Hawking radiation of light primordial black holes (PBH) in presence of a non-zero chemical potential, originating from the space-time curvature. Considering the PBHs are described by a Reissner-Nordstr\"{o}m metric, and are produced in a radiation dominated Universe, we show, it is possible to simultaneously explain the matter-antimatter asymmetry along with right DM abundance satisfying bounds from big bang nucleosynthesis, cosmic microwave background and gravitational wave energy density due to PBH density fluctuation. We also obtain the parameter space beyond the semiclassical approximation, taking into account the quantum effects on charged PBH dynamics due to memory burden.   
}
\begin{flushright}
\small{HRI-RECAPP-2024-08}
\end{flushright}
\maketitle

\section{Introduction}
\label{sec:intro}
Since the pioneering work by Stephen Hawking~\cite{Hawking:1974rv, Hawking:1975vcx}, primordial black holes (PBHs) have captivated astrophysicists and cosmologists. There are many mechanisms for PBH formation, for example, bubble collisions~\cite{Hawking:1982ga,Kodama:1982sf} or collapse of scalar fields~\cite{Green:2014faa,Cotner:2018vug}. It is usually assumed that the formation of PBHs happened from some time during inflation to the radiation-dominated epoch~\cite{Leach:2000yw}. PBHs with lifetimes longer than the Universe's age, specifically those with $\mbh \gtrsim 10^{15}$ g, could account for dark matter (DM) within the mass window: $10^{17}\lesssim\mbh/\text{g}\lesssim 10^{23}$~\cite{Carr:2020gox, Green:2020jor}. These objects might also source gravitational waves (GW)~\cite{Bird:2016dcv, DeLuca:2020qqa, NANOGrav:2023hvm} and explain other cosmic events~\cite{Barack:2018yly, Sasaki:2018dmp}. Conversely, lighter PBHs ($\mbh \lesssim 10^9$ g) may have shaped the early Universe by dominating Universe's energy density before evaporation. Their decay could produce particles within and beyond the Standard Model (SM), and even contribute to dark matter production~\cite{Morrison:2018xla, Gondolo:2020uqv, Bernal:2020bjf, Green:1999yh, Khlopov:2004tn, Dai:2009hx, Allahverdi:2017sks, Lennon:2017tqq, Hooper:2019gtx, Chaudhuri:2020wjo, Masina:2020xhk, Baldes:2020nuv, Bernal:2020ili, Bernal:2020kse, Lacki:2010zf, Boucenna:2017ghj, Adamek:2019gns, Carr:2020mqm, Masina:2021zpu, Bernal:2021bbv, Bernal:2021yyb, Samanta:2021mdm, Sandick:2021gew, Cheek:2021cfe, Cheek:2021odj, Barman:2021ost, Borah:2022iym,Chen:2023lnj,Chen:2023tzd,Kim:2023ixo,Gehrman:2023qjn,Calza:2023rjt,Coleppa:2022pnf,Chaudhuri:2023aiv}, baryon asymmetry~\cite{Baumann:2007yr, Fujita:2014hha, Morrison:2018xla, Hooper:2020otu, Perez-Gonzalez:2020vnz, Datta:2020bht, JyotiDas:2021shi, Smyth:2021lkn, Barman:2021ost, Bernal:2022pue, Ambrosone:2021lsx,Calabrese:2023key,Calabrese:2023bxz,Gehrman:2022imk,Gehrman:2023esa,Schmitz:2023pfy}, and cogenesis~\cite{Fujita:2014hha, Morrison:2018xla, Hooper:2019gtx, Lunardini:2019zob, Masina:2020xhk, Hooper:2020otu, Datta:2020bht, JyotiDas:2021shi, Schiavone:2021imu, Bernal:2021yyb, Bernal:2021bbv, Bernal:2022swt,Barman:2022pdo,Borah:2024qyo,Chianese:2024nyw}. These studies assume PBHs remain classical throughout their lifetime, validating a semiclassical approach. However, Hawking's framework loses accuracy when PBHs evaporate half their initial mass, a phenomenon known as the memory burden~\cite{Dvali:2018xpy, Dvali:2020wft}, that arises from quantum information retained by the system, creating a backreaction that stabilizes the PBH halfway through its decay. Recent studies have explored memory burden's implications for DM, baryon asymmetry, and primordial GW signatures, see, for example, Refs.~\cite{Balaji:2024hpu, Haque:2024eyh,Barman:2024iht,Borah:2024bcr,Thoss:2024hsr,Dvali:2024hsb,Alexandre:2024nuo, Barman:2024ufm,Bhaumik:2024qzd,Basumatary:2024uwo,Athron:2024fcj,Loc:2024qbz}.

The origin of the Universe's baryon asymmetry and the nature of cosmological DM are among the most profound mysteries in modern physics. These challenges lie at the crossroads of cosmology and particle physics, inspiring extensive efforts in both fields. The well-known and elegant method of generating a non-zero baryon asymmetry is via leptogenesis~\cite{Yanagida:1979as}, that requires physics beyond the SM (the right handed neutrinos), and can be sourced by PBH evaporation~\cite{Carr:1976zz}. The same is also true for a viable particle DM candidate. However, the idea of generation of baryon asymmetry from PBH evaporation is rather a more natural one. The Hawking radiation itself could be asymmetric due to a dynamically generated chemical potential near the surface of the black hole. This is a purely gravitational phenomenon and can well be dubbed as {\it gravitational baryogenesis}.  In Ref.~\cite{Davoudiasl:2004gf}, a mechanism of gravitational baryogenesis was prescribed whereby one could generate the adequate baryon asymmetry by introducing a quantum gravity inspired\footnote{{ As  mentioned in Ref.~\cite{Lambiase:2013haa}, such an effective interaction is expected in a low energy effective theory of quantum gravity or Super gravity theories (more specifically it can be obtained in supergravity theories from a higher dimensional operator in the K\"{a}hler potential~\cite{Davoudiasl:2004gf}). While the origin of the interaction in Eq.~\eqref{eq:action1} was originally envisioned to arise from quantum gravity, it was
later argued that an effective interaction of this form is
a generic outcome in models that feature high-scale CP violation in a curved spacetime background, which can result in a cutoff scale $M_\star\ll M_P$~\cite{McDonald:2014yfg,McDonald:2015iwt,McDonald:2020ghc,Samanta:2020cdk}. It was shown that the same interaction is generated dynamically at two loops in the type-I seesaw mechanism, when the seesaw Lagrangian is minimally coupled to the gravitational background as a direct consequence of CP violation in the RHN Yukawa sector. One may also replace $\mathcal{R}$ in Eq.~\eqref{eq:action1} with $f(\mathcal{R})$~\cite{Li:2004hh,Lambiase:2006dq}, motivating more generalized gravity theories, or even modified theories of gravity~\cite{Oikonomou:2016jjh,Odintsov:2016hgc}}.} coupling between the Ricci scalar $\mathcal{R}$, that encodes the information of the background metric, and the $B-L$ current $j_{B-L}^\mu$,
\begin{align}\label{eq:action1}
&\mathcal{S}_{B-L}= \int d^4x\,\sqrt{-g}\,\lambda\,\frac{\partial_\mu\,\mathcal{R}}{M_P^2}\,j^\mu_{B-L}\,. 
\end{align}
Due to the expansion of the universe, $\dot{\mathcal{R}}\neq 0$, so it generates a chemical potential for the $B-L$ current\footnote{Similar studies have also been done in Refs.~\cite{Modak:2014xza,Lima:2016cbh}.} in such a way the Hamiltonian density becomes corrected by the term 
\begin{align}
    \mu_{B-L}\,n_{B-L}&=\mathcal{L}_{B-L}=\lambda\,\frac{\partial_0\,\mathcal{R}}{M_P^2}\,j^0_{B-L}\,,
\end{align}
where the fact that in the Friedmann Lemaitre Robertson Walker (FLRW) background the current $j^\mu_{B-L}$ is homogeneous, has been invoked. The action Eq.~(\ref{eq:action1}) thus gives rise to asymmetric Hawking radiation that produces a baryon number. Specifically, it allows for the production of a non-vanishing baryon-lepton charge that sources the asymmetry. In Ref.~\cite{Hook:2014mla}, it has been shown how asymmetric Hawking radiation can be initiated due to a dynamically generated baryonic chemical potential at the horizon. As shown in Ref.~\cite{Boudon:2020qpo}, in order to satisfy the observed baryon asymmetry, together with PBHs as DM, the coupling $\lambda$ needs to be as large as $\mathcal{O}(10^{100})$. Such value of the dimensionless coupling albeit seems ``unusual" but in principle, given our ignorance of a full quantum gravity theory, is perfectly acceptable~\cite{Hook:2014mla}. In Refs.~\cite{Hamada:2016jnq,Smyth:2021lkn}, a different action has been considered, where a non-zero Kretschmann scalar is responsible for the generation of a non-vanishing chemical potential. Here also the PBHs that could have produced the observed baryon asymmetry through gravitational baryogenesis, have simultaneously accounted for the DM abundance.

Motivated from these, in this work we have considered Eq.~\eqref{eq:action1} as our starting point to generate the  asymmetry in the visible sector, via asymmetric Hawking evaporation from a population of PBHs that evaporate before the onset of primordial nucleosynthesis. The PBHs therefore are essentially light and can not account for the observed DM abundance of the Universe. The DM, on the other hand, is generated via asymmetric Hawking evaporation triggered by an action similar to Eq.~\eqref{eq:action1}, but in the dark sector. We show in such a case it is indeed possible to generate asymmetries both in the visible and in the dark sector that can account for the baryon-DM coincidence problem~\cite{Planck:2018vyg}. We further take the effect of memory burden into account and explore the viable parameter space that satisfies both the abundance.  

The paper is organized as follows. In Sec.~\ref{sec:dynamics} we discuss the generalities of charged PBH evaporation, followed by the mechanism of generation of baryon asymmetry. The production of asymmetric DM is discussed in Sec.~\ref{sec:DM}. In Sec.~\ref{sec:memory} we discuss the aftermath of memory-burden effect. All relevant numerical results are provided in Sec.~\ref{sec:result}. Finally, we conclude in Sec.~\ref{sec:concl}.
\section{Dynamics of a charged BH}
\label{sec:dynamics}
We consider the Reissner-Nordstr\"{o}m (RN) metric, which is a solution to Einstein’s field equations that describes the spacetime around a spherically symmetric non-rotating body with mass $M$ and an electric charge $Q$. Other than spherical symmetry we also have the assumption that the space is empty from matter (there is only an electromagnetic field). 
The RN metric can be written as
\begin{align}
& ds^2 = -\frac{\Delta }{r^2}\,dt^2+\frac{r^2}{\Delta}\,dr^2+r^2\,d\Omega^2\,,    
\end{align}
where
\begin{align}
\Delta = r^2-2\,G\mbh\,r+G\,Q^2=(r-r_+)\,(r-r_-)\,, 
\end{align}
with 
\begin{align}
r_\pm = \mbh G\pm\sqrt{\mbh^2\,G^2-G\,Q^2}\,. 
\end{align}
We denote $M_P=1/\sqrt{8\pi\,G}$ as the reduced Planck mass. Clearly, when $Q\to 0$, the metric should approach the Schwarzschild metric. For $\mbh<|Q|/\sqrt{G}$, no horizon exists and the singularity at $r = 0$ is naked. On the other extreme, for $\mbh>|Q|/\sqrt{G}$, $\Delta=0$ or $r_\pm$, and the metric becomes singular. However, this is just a coordinate singularity. For $\mbh=|Q|/\sqrt{G}$, we get an extremal BH, where the inner and outer horizons merge into one. The temperature of the charged BH is given by
\begin{align}\label{eq:TBH}
\Tbh=4M_P^2\,\frac{\sqrt{\mbh^2-Q^2/G}}{\left(\mbh+\sqrt{\mbh^2-Q^2/G}\right)^2}\,,    
\end{align}
with corresponding entropy,
\begin{align}\label{eq:SBH}
& S_{\rm BH}\equiv\frac{4\pi\,r_+^2}{4G} =\frac{1}{8}\,\left[\left(\frac{\mbh}{M_P}\right)+\sqrt{\left(\frac{\mbh}{M_P}\right)^2-8\,\pi\,Q^2}\right]^2\,.    
\end{align}
Note that, for $Q\to0$, we obtain the Hawking temperature for a Schwarzschild BH: $\Tbh=M_P^2/\mbh$ and $S_{\rm BH}=(1/2)\,\left(\mbh/M_P\right)^2$.

Since the BH has an electric charge, the ensuing electric potential plays the role of a chemical potential $\mu$ at the horizon surface,  in such a way the BH thermally evaporates into charged particles with charge $q_i$ with a non zero chemical potential $\mu_i$.  
Now, the energy of loss per unit area per unit time due to Hawking radiation is given by
\begin{align}
 \frac{dE}{dx^2\,dt}&=-\sum_i\frac{g_i}{4}\int\,\frac{d^3p}{(2\pi)^3}\,\frac{p}{e^{\left(p+\mu_i\right)/\Tbh}\pm1}\nonumber\\
&=-\sum_{i=f}\frac{g_i}{4}\,\Tbh^4\,f\left(\frac{\mu_i}{\Tbh}\right)-\sum_{i=b}\frac{g_i}{4}\,\Tbh^4\,b\left(\frac{\mu_i}{\Tbh}\right)\,,   
\end{align}  
where $g_i$ is the degree of freedom and $\mu_i$ is the chemical potential associated with evaporated species $i$. The functions, $f(x)=\left(-3/\pi^2\right)\,\text{Li}_4\left[-e^{-x}\right]$ and $b(x)=\left(3/\pi^2\right)\,\text{Li}_4\left[e^x\right]$, with $\text{Li}_a[z]=\sum_{k=1}^\infty z^k/k^a$ being a polylogarithmic function. The total energy lost per unit time is obtained by multiplying with the area of the event horizon $4\pi\,r_+^2$, \begin{align}
\frac{dE}{dt}\simeq-\frac{G\sqrt{\mbh^2-Q^2/G}}{2}\,\Tbh^3\,\left(\sum_{i=f} g_i\,f(\mu_i)+\sum_{i=b} g_i\,b(\mu_i)\right)\,,  
\end{align}
where $\mu_i\ll \Tbh$. This can be further simplified to,
\begin{align}\label{eq:dEdt}
& \frac{dE}{dt}\simeq-\frac{M_P^4}{16\pi\mbh^2}\,\left(\sum_{i=f} g_i\,f(\mu_i)+\sum_{i=b} g_i\,b(\mu_i)\right)=-\epsilon\,\frac{M_P^4}{\mbh^2}\,,
\end{align}
assuming $\mbh\gg Q/\sqrt G$ and considering the BH to be a perfect blackbody with $\epsilon=27/4\times g_{\star,H}(\Tbh)\,\pi/(480)$, where the factor 27/4 accounts for the graybody factor~\cite{PhysRevD.41.3052,Auffinger:2020afu,Masina:2021zpu,Cheek:2021odj}. Here,  
\begin{equation}
g_{\star,H}(T_\text{BH})\equiv\sum_i\omega_i\,g_{i,H}\,; g_{i,H}=
    \begin{cases}
        1.82
        &\text{for }s=0\,,\\
        1.0
        &\text{for }s=1/2\,,\\
        0.41
        &\text{for }s=1\,,\\
        0.05
        &\text{for }s=2\,,\\
    \end{cases}
    \label{eq:gstTBH}
\end{equation}
with $\omega_i=2\,s_i+1$ for massive particles of spin $s_i$, $\omega_i=2$ for massless species with $s_i>0$ and $\omega_i=1$ for $s_i=0$. At temperatures $T_\text{BH}\gg T_\text{EW}\simeq 160$ GeV, BH evaporation emits the full SM particle spectrum according to their $g_{\star,H}$ weights, while at temperatures below the MeV scale, only photons and neutrinos are emitted. For $T_\text{BH}\gg 100$ GeV, the particle content of the SM corresponds to $g_{\star,H}(T_\text{BH})\simeq 106.75$. Assuming $g_{\star,H}(\Tbh)$ remians approximately constant over the BH lifetime, we can integrate Eq.~\eqref{eq:dEdt} to obtain the BH mass evolution as
\begin{align}\label{eq:mbh}
& \mbh(t)\simeq \Min\,\left[1-3\,\epsilon\,\frac{M_P^4\,\tin}{\Min^3}\,\left(\frac{t}{\tin}-1\right)\right]^{1/3}\,,    
\end{align}
where $\mbh(t=\tin)=\Min$ is the BH mass at formation. This provides the BH lifetime as, 
\begin{align}\label{eq:tau}
& \tau = \tin+\frac{\Min^3}{3\,\epsilon\,M_P^4}\,.    
\end{align}
If $q_i$ is the electric charge of a species $i$ emitted during BH evaporation, the rate of electric charge loss due to BH evaporation is given by,
\begin{align}\label{eq:dQdt1}
\frac{dQ}{dt}&=4\pi\,r_+^2\,\sum\frac{g_i\,q_i}{4}\,\left(n_i-\bar n_i\right)\nonumber\\
&= 4\pi\,r_+^2\,\sum\frac{g_i\,q_i}{4}\,\int\,\frac{d^3p}{(2\pi)^3}\,\left(\frac{1}{\exp\left[\frac{p+\mu_i}{\Tbh}\right]+1}-\frac{1}{\exp\left[\frac{p-\mu_i}{\Tbh}\right]+1}\frac{}{}\right)  
\nonumber\\&
\simeq 
-4\pi\,r_+^2\,\sum\frac{g_i\,q_i}{4}\,\left(\frac{\mu_i\,\Tbh^2}{6}+\frac{\mu_i^3}{6\pi^2}+\mathcal{O}\left[\frac{\mu_i}{\Tbh}\right]^2\right)\,,
\end{align}
which leads to \begin{align}\label{eq:dQdt}
&\frac{dQ}{dt}
\simeq -Q\,\sum_i\frac{g_i\,q_i^2}{24}\,T_{\rm BH}+\mathcal{O}[Q^3]\,,
\end{align}
where $\mu_i=q_i\,Q/r_+$ is the chemical potential that originates from the BH charge and we have considered $\mu_i/T\ll 1$. The above equation shows 
\begin{align}\label{eq:Qem}
& Q(t)\simeq\Qin\exp\left[-\sum\frac{g_i\,q_i^2}{24\,\pi}\,\frac{M_P^2\,t}{\Min}\right]\,, 
\end{align}
where we have considered $M_P^4\,t/\Min^3\ll 1$ and used Eq.~\eqref{eq:mbh}. Considering contribution solely from the SM fermions, $\sum g_i\,q_i^2=36\times (2/3)^2+36\times (-1/3)^2+12\times (-1)^2=32$. Here $\Qin=Q(\tin)$ is the initial BH charge. Thus, the electric charge of the BH is lost exponentially with time. 
\subsection{Baryon asymmetry from a charged BH}
Following ``no-hair theorem", a BH is fully characterized by just three observable classical parameters: mass, electric charge, and angular momentum. It implies that once a BH forms, any other information about the matter that collapsed to form it is lost (or ``invisible" to an external observer). As a result, a BH does not retain any information about global quantum numbers, e.g., baryon number. Thus, the non-existence of the BH's global charge (due to the no-hair theorem) implies that losing particles or radiation with that charge does not change the BH's fundamental parameters. Therefore, if continuous global symmetries such as $B-L$ existed, when a charged particle becomes trapped inside a black hole, it would be impossible to detect this charge from outside the event horizon. As a result, the charge would seem to be `erased,' contradicting the principle of charge conservation~\cite{Kallosh:1995hi}.

In presence of an explicit chemical potential at the BH horizon with radius $r_{\rm BH}=2\,G\,\mbh$, we can follow the exact same approach as before that leads to a global $B-L$ charge evolution given by,
\begin{align}\label{eq:BL}
& \frac{dQ_{B-L}}{dt}\simeq4\,\pi\,r_{\rm BH}^2\times\frac{\Tbh^2}{24} \sum g_i\,\mu_i\,q_i^{bl}
=\sum \frac{g_i\,\mu_i\,q_i^{bl}}{96\pi}
\,,   
\end{align}
where we have dropped $\mu_i^3$ relative to $\mu_i\,T^2$, assuming $\mu_i\ll T$, and also $\mbh\gg Q/\sqrt G$. Also, note that the above equation certainly does not follow charge conservation. Here $q_i^{bl}$ is the $B-L$ charge of the emitted particles, while the sum is over particles only. Further, in Eq.~\eqref{eq:BL} we have used the Fermi–Dirac distribution exclusively,  since in the SM only fermions carry baryon number. If one assumed that bosons could carry baryon number then one should also use the Bose–Einstein distribution. 

Because the expansion of the Universe, the Ricci scalar $\mathcal R$ induces a chemical potential given by
$\mu_i=\lambda\,q_i^{bl}\,\dot{\mathcal{R}} /(8\pi\,M_P^2)$. Then, the $Q_{B-L}$ loss rate becomes 
\begin{align}
& \frac{dQ_{B-L}}{dt}=\lambda\sum \frac{g_i\,\left(q_i^{bl}\right)^2}{96\pi}\,\frac{\dot{\mathcal{R}}}{8\pi\,M_P^2}\,,    
\end{align}
where the sum is over only particles. Now, The time derivative of the Ricci scalar in a FLRW background~\cite{Hook:2014mla},
\begin{align}
\dot{\mathcal{R}}=-9\,(1+w)\,(1-3w)\,H^3\,,    
\end{align}
where $w$ denotes the background equation of state. To the lowest order, no net asymmetry is generated during the de Sitter $(w=-1)$ and during the radiation dominated $(w=1/3)$ phase, however this is not true for matter dominated (MD) epoch, for which $w=0$. At the one loop level, including quantum correction one finds~\cite{Kajantie:2002wa,Hook:2014mla,Modak:2014xza}
\begin{align}
& 1-3w=\frac{5}{6\pi^2}\,\left(\frac{g_3^2}{4\pi}\right)^2\,\frac{\left(N_c+5N_f/4\right)\,\left(11N_c/3-2N_f/3\right)}{2+(7/2)\,\left(N_c\,N_f/(N_c^2-1)\right)}\,,    
\end{align}
where $N_c$ is the number of colors, $N_f$ is the
number of flavors and $\alpha_3\equiv g_3^2/(4\pi)\approx 0.118$~\cite{ParticleDataGroup:2024cfk} is the $SU(3)$ fine structure constant. Within the SM, $N_c=3,\,N_f=6$, and $1-3w\simeq 8\times 10^{-3}$, showing that $1-3w$ has a non-zero value. Further, for SM, $\sum g_i\,(q_i^{bl})^2=36\times (1/3)^2+9\times(-1)^2=13$. Thus,
\begin{align}\label{eq:B-L}
& \frac{dQ_{B-L}}{dt}=-\frac{245\,\lambda}{5056\,\pi^4}\,\frac{\alpha_3^2}{M_P^2}\times\sum g_i\,\left(q_i^{bl}\right)^2
\begin{cases}
\left(\frac{1}{2t}\right)^3 & \text{for RD}\,,
\\[10pt]
\frac{237\,\pi^2}{980\,\alpha_3^2}\,\left(\frac{2}{3t}\right)^3 & \text{for MD}\,,
\end{cases}
\end{align}
where in the second case, we have $w=0$. One can then integrate the above expressions till the BH lifetime [cf.Eq.~\eqref{eq:tau}] and obtain
\begin{align}\label{eq:QBL}
& Q_{B-L}(\tau)\simeq -\frac{\lambda}{M_P^2}\,\sum g_i\,\left(q_i^{bl}\right)^2\times
\begin{cases}
\frac{245\,\alpha_3^2}{80896\,\pi^4}\,\frac{1}{\tin^2} & \text{for RD}\,,
\\[10pt]
\frac{1}{576\,\pi^2}\,\frac{1}{\tin^2} & \text{for MD}\,,
\end{cases}
\end{align}
at the end of BH evaporation, where $\tau\approx\tev\gg\tin$ is assumed. Since, $Q_{B-L}$ is equivalent to the $B-L$ number, we can write the final asymmetry as
\begin{align}
& Y_B^0=\frac{n_B}{s}\Big|_0=Q_{B-L}\,a_{\rm sph}\,\frac{n_{\rm BH}}{s}\Big|_{\tev\approx\tau}=Q_{B-L}\,a_{\rm sph}\,Y_{\rm BH}(\tev)\,,  
\end{align}
where $a_{\rm sph}=28/79$ is the spahelron conversion factor and $Y_B^0\simeq 8.7\times 10^{-11}$ is the observed baryon asymmetry today~\cite{Planck:2018jri}. Here we have assumed that there is no entropy injection in the thermal bath once the BH evaporation is complete. Considering BH is produced in a radiation dominated Universe,
\begin{align}\label{eq:YBH1}
& Y_{\rm BH}(\Tev)\Big|_{\beta < \beta_c}\simeq\frac{3\,\sqrt{3}\beta}{2}\,\mathcal{F}(\gs,\,\gss)\,\left(\frac{10\,\gamma^2}{\gs(\Tin)}\right)^{1/4}\,\left(\frac{M_P}{\Min}\right)^{3/2}\,, 
\end{align}
where 
\begin{align}
\mathcal{F}(\gs,\,\gss) =\left(\frac{\gs(\Tev)}{\gs(\Tin)}\right)^{3/4}\,\left(\frac{\gs(\Tin)}{\gss(\Tev)}\right)\,.
\end{align}
where $\beta=\rho_{\rm BH}(\tin)/\rho_R(\tin)$, with 
\begin{align}\label{eq:betac}
\beta_c=\left(\frac{\gs(\Tev)}{\gs(\Tin)}\right)^{1/4}\,\frac{\Tev}{\Tin}=\sqrt{\frac{\tin}{\tau}}\,,    
\end{align}
being the critical $\beta$-values above which we always have BH domination. For BHs formed in RD,
\begin{align}\label{eq:Min}
& \Min=\gamma\,\frac{4\pi}{3}\,\rho_R\,\left(\frac{1}{H(\tin)}\right)^3\implies\tin=\frac{\Min}{8\pi\gamma\,M_P^2}\,,    
\end{align}
where $\gamma\simeq 0.2$, is a numerical factor that parameterizes the efficiency of the collapse to form PBHs and the Hubble parameter $H(t)=1/(2t)$. One can then obtain corresponding temperature at the epoch of BH formation,
\begin{align}\label{eq:Tin}
& \Tin=\left(12\,\gamma\,\sqrt{\frac{10}{\gs(\Tin)}}\,M_P^2\right)^{1/2}\,\sqrt{\frac{M_P}{\Min}}\,.    
\end{align}
Therefore, the final yield for BH evaporation during radiation domination
\begin{align}
& Y_B^0\simeq 3\sqrt{3}\,\beta\,\left|Q_{B-L}\right|\,a_{\rm sph}\,\left(\frac{5\gamma^2}{8}\right)^{1/4}\,\left[\left(\frac{\gs(\Tin)}{\gss(\Tev)}\right)\,\left(\frac{\gss(\Tev)}{\gss(\Tin)}\right)^4\right]^{1/4}\,\left(\frac{M_P}{\Min}\right)^2\,, 
\end{align}
where we have neglected the time-dependence of $\mbh$. 
For BH evaporation during BH domination, using the Friedmann equation we find,
\begin{align}
& Y_{\rm BH}(\Tev)=\frac{n_{\rm BH}(\Tev)}{s(\Tev)}=\frac{30}{\pi^2\,\gss(\Tev)}\,\frac{M_P^2}{\tau^2\,\Tev^3\,\Min}\,,  
\end{align}
where we have ignored the evolution of the BH mass. Considering the Universe is still radiation dominated at $t=\tev$, we obtain
\begin{align}\label{eq:Tevap}
\Tev=\left(\frac{45}{(2\pi^2)\,\gs(\Tev)}\right)^{1/4}\,\sqrt{\frac{M_P}{\tau}}\,,    
\end{align}
which leads to
\begin{align}\label{eq:YBH2}
& Y_{\rm BH}(\overline{T}_{\rm ev})\Big|_{\beta > \beta_c}=\left(\frac{405}{32}\right)^{1/4}\,\sqrt{\frac{3\epsilon}{\pi}}\,\left(\frac{\gs(\overline{T}_{\rm ev})^3}{\gss(\overline{T}_{\rm ev})^4}\right)^{1/4}\,\left(\frac{M_P}{\Min}\right)^{5/2}\,.  
\end{align}
It is important to note here, if the PBH component dominates at some point the total energy density of the
universe, the SM temperature just after the complete evaporation of PBHs is
\begin{align}
& \overline{T}_{\rm ev}=\frac{2}{\sqrt{3}}\,\Tev\,.    
\end{align}
The final asymmetry in these two cases then reads,
\begin{align}\label{eq:YB0}
& Y_B^0\simeq\sqrt{3}\left|Q_{B-L}\right|\,a_{\rm sph}\times
\begin{cases}
\frac{3\sqrt{3}\,\beta}{2}\,\left(\frac{10\,\gamma^2}{\gs(\Tin)}\right)^{1/4}\,\left[\left(\frac{\gs(\Tev)}{\gs(\Tin)}\right)^{3/4}\,\frac{\gs(\Tin)}{\gss(\Tev)}\right]\,\left(\frac{M_P}{\Min}\right)^{3/2}
&\beta<\beta_c\,,
\\[10pt]
\left(\frac{405}{32}\right)^{1/4}\,\sqrt{\frac{3\epsilon}{\pi}}\,\left(\frac{\gs(\overline{T}_{\rm ev})^3}{\gss(\overline{T}_{\rm ev})^4}\right)^{1/4}\,\left(\frac{M_P}{\Min}\right)^{5/2} &\beta>\beta_c\,,
\end{cases}
\end{align}
where in the second case we note the $\beta$-independence and $|Q_{B-L}|$ is given by Eq.~\eqref{eq:QBL}.  We notice that the yield has a smooth transition between the two eras, when $\beta\to\beta_c$ and $\Tev=\overline{T}_{\rm ev}$. On substituting Eq.~\eqref{eq:B-L} we find, in order to produce the observed baryon asymmetry, one needs to have 
\begin{align}\label{eq:lam-YB}
& \lambda\simeq
\begin{cases}
1.2\times 10^{30}\times\left(\frac{10^{-10}}{\beta}\right)\,\left(\frac{\Min}{100\,\text{g}}\right)^{7/2} & \beta<\beta_c\,,
\\[10pt]
8.2\times 10^{26}\times\left(\frac{\Min}{100\,\text{g}}\right)^{9/2} & \beta>\beta_c\,,
\end{cases}
\end{align}
considering the number of relativistic degrees of freedom to be fixed at 106. At this stage, we have three independent parameters in our theory
\begin{align}
\{\Min,\,\beta,\,\lambda\}\,.   
\end{align}

The production of gravitational waves (GW) induced by large-scale density perturbations underlain by PBHs could lead to a backreaction
problem. However, demanding that the energy contained in GWs never overtakes the
one of the background Universe one can obtain an upper bound on $\beta$ as~\cite{Domenech:2020ssp},
\begin{align}\label{eq:betaGW}  
\beta<1.1\times 10^{-6}\,\left(\frac{g_{\star,H}(\Tbh)}{108}\right)^{17/48}\,\left(\frac{\gs(\Tev)}{106.75}\right)^{-1/3}\,\left(\frac{\Min}{10^4\,{\rm g}}\
\right)^{-17/24}\,.
\end{align}
Since PBH evaporation produces all particles, hence it poses a potential threat to the successful predictions of the Big Bang Nucleosynthesis (BBN) by introducing extra relativistic degrees of freedom around BBN. Thus, we PBHs need to fully evaporate
before the onset of BBN, or, in other words, $\Tev>T_{\rm BBN}\simeq 4$ MeV~\cite{Sarkar:1995dd, Kawasaki:2000en,Hannestad:2004px, DeBernardis:2008zz, deSalas:2015glj,Hasegawa:2019jsa} which translates into an upper bound on the initial PBH mass. A lower bound on $\Min$ can be set once the upper bound on the inflationary scale is taken into account~\cite{Planck:2018vyg}: $H_I\lesssim 2\times 10^{-5}\,M_P$.
Therefore, the viable PBH mass window becomes: $\Min\in\left(0.1,\,10^8\right)$ g. However, in order for the sphaleron process to take place efficiently, we need $\Tev>T_{\rm EW}\simeq 160$ GeV~\cite{Fujita:2014hha}. This results in a tighter constraint on the upper bound bound on BH mass compared to BBN, discarding $\Min\gtrsim 10^5$ g.
\section{Asymmetric dark matter from a charged BH}
\label{sec:DM}
We now explore the possibility that the DM abundance is entirely due to the asymmetric component of a DM species, generated gravitationally because of a non-zero chemical potential generated in the dark sector. To achieve that, we write a pure gravitational interaction similar to the action in Eq.~\eqref{eq:action1} as,
\begin{align}\label{eq:action-dm}
&\mathcal{S}_{\rm DM}= \int d^4x\,\sqrt{-g}\,\lambda_{\rm DM}\,\frac{\partial_\mu\,\mathcal{R}}{M_P^2}\,j^\mu_{\rm DM}\,, 
\end{align}
with $\mu_{\rm DM}=\lambda_{\rm DM}\,q_{\rm DM}\,\mathcal{\dot{R}}/(8\pi\,M_P^2)$, where $\qdm$ is the charge (equivalent to $q^{bl}$ in the visible sector) carried by each DM particle and a summation over the particles is implied. Here $j_{\rm DM}^\mu$ is a current in the DM sector that generates an asymmetry within the dark sector only and $\ldm$ is a free parameter that we shall constraint by demanding right abundance for the DM. Proceeding same as before, we can write down the time-evolution of net DM charge asymmetry as,  
\begin{align}
& \frac{dQ_{\rm DM}}{dt}=\frac{3\,\Ldm\,g_{\rm DM}}{256\,\pi^2\,M_P^2}\,\,(1+w)\,(3w-1)\,H^3\,,  
\end{align}
which is similar to Eq.~\eqref{eq:BL}, with $\Lambda_{\rm DM}=\ldm\,\qdm^2$. Here $g_{\rm DM}$ is the degrees of freedom of the corresponding DM candidate that depends on the DM spin. We integrate the above equation from $\tin$ till $\tev$ to obtain the total dark charge generated,
\begin{align}
Q_{\rm DM}(\tau)\simeq-\frac{\Ldm\,g_{\rm DM}}{M_P^2\,\pi^2}\,\frac{1}{\tin^2}
\begin{cases}
\frac{135}{161792\,\pi^4} & \text{for RD}
\\[10pt]
\frac{1}{576} & \text{for MD}\,,
\end{cases}
\end{align}
where $\tev\gg\tin$ is assumed. Again, since this is equivalent to $n_{\rm DM}-n_{\overline{\rm DM}}$, i.e., difference between DM and anti-DM, we can determine the final DM yield as,
\begin{align}\label{eq:YDM}
& Y_{\rm DM}^0=\left|Q_{\rm DM}\right|\,\frac{n_{\rm BH}}{s}\Big|_{\tev\approx\tau}
\nonumber\\&    
\simeq\left|\Qdm\right|\times
\begin{cases}
\frac{3\sqrt{3}\,\beta}{2}\,\left(\frac{10\,\gamma^2}{\gs(\Tin)}\right)^{1/4}\,\left[\left(\frac{\gs(\Tev)}{\gs(\Tin)}\right)^{3/4}\,\frac{\gs(\Tin)}{\gss(\Tev)}\right]\,\left(\frac{M_P}{\Min}\right)^{3/2}
&\beta<\beta_c\,,
\\[10pt]
\left(\frac{405}{32}\right)^{1/4}\,\sqrt{\frac{3\epsilon}{\pi}}\,\left(\frac{\gs(\overline{T}_{\rm ev})^3}{\gss(\overline{T}_{\rm ev})^4}\right)^{1/4}\,\left(\frac{M_P}{\Min}\right)^{5/2} &\beta>\beta_c\,,
\end{cases}
\,.
\end{align}

To fit the whole observed DM relic density, it is required that
\begin{equation} \label{eq:obsyield}
    Y^0_{\rm DM}\, \mdm = \Omega_{\rm DM} h^2 \, \frac{1}{s_0}\,\frac{\rho_c}{h^2} \simeq 4.3 \times 10^{-10} {\rm GeV}\,,
\end{equation}
where $\rho_c \simeq 1.05 \times 10^{-5}\, h^2$~GeV/cm$^3$ is the critical energy density, $s_0\simeq 2.69 \times 10^3$~cm$^{-3}$ the present entropy density~\cite{Planck:2018vyg}, and $\Omega_{\rm DM} h^2 \simeq 0.12$ the observed abundance of DM relics~\cite{Planck:2018vyg}. On top of $\beta$ and $\Min$, we now have two more independent parameters for a given DM spin, namely the DM mass $\mdm$ and $\Ldm$. In order to to saturate the observed DM relic abundance, considering a scalar DM candidate $(g_{\rm DM}=1)$, we find
\begin{align}\label{eq:lam-DM}
& \Ldm\simeq \frac{\mdm}{10^5\,\text{GeV}}\,
\begin{cases}
1.4\times 10^{36}\times\left(\frac{10^{-10}}{\beta}\right)\,\left(\frac{\Min}{100\,\text{g}}\right)^{7/2} & \beta<\beta_c\,,
\\[10pt]
9.4\times 10^{32}\times\left(\frac{\Min}{100\,\text{g}}\right)^{9/2} & \beta>\beta_c\,,
\end{cases}
\end{align}
considering the number of relativistic degrees of freedom to be fixed at 106. From Eqs.~\eqref{eq:lam-YB} and \eqref{eq:lam-DM}, we note, in order to generate adequate asymmetry in the visible and in the dark sector, the couplings need to be $\gtrsim\mathcal{O}(10^{20})$\footnote{ {Although the coupling are $\sim\mathcal{O}(10^{20})$, such interactions could still remain undetectable in usual laboratory experiments because of $M_P^2$ suppression}.}.

If DM is to account for the full relic abundance, it must be sufficiently cold to avoid disrupting the formation of large-scale structures. A DM candidate that is either part of the thermal bath or produced from it should have a mass greater than a few keV to ensure the required free-streaming properties, as constrained by the Lyman (Ly)-$\alpha$ flux-power spectrum~\cite{Irsic:2017ixq,Ballesteros:2020adh}. However, such a low DM mass is tightly constrained by the free-streaming behavior of ultra-relativistic DM species. As shown in Refs.~\cite{Fujita:2014hha,Masina:2020xhk}, one can derive a lower bound on the DM mass originating from PBH evaporation as, 
\begin{align}\label{eq:wdm}
& \mdm\gtrsim 10^4\,\langle E_{\rm DM}(t_{\rm eq})\rangle\,,    
\end{align}
where $\langle...\rangle$ stands for the average over PBH temperature and we consider the bound on WDM from Ly-$\alpha$ to be 3.5 keV~\cite{PhysRevD.96.023522}. The average kinetic energy (KE) of the DM particles then reads
\begin{align}\label{eq:EDM}
\langle E_{\rm DM}(t_{\rm eq})\approx\langle E_{\rm DM}(t_{\rm ev}^q)\rangle\,\frac{a_{\rm ev}}{a_{\rm eq}}\simeq \delta\times\Tbh\,\frac{T_{\rm eq}}{T_{\rm ev}}\,\left[\frac{\gss(T_{\rm eq})}{\gss(T_{\rm ev})}\right]^{1/3}\,,    
\end{align}
with $T_{\rm eq}\simeq 0.75$ eV being the temperature at the epoch of late matter-radiation equality (MRE) while $\delta$ is a parameter of order 1. As we will see, this will appear as an important bound in constraining the allowed parameter space. 
\section{A charged BH with memory burden}
\label{sec:memory}
So far we have discussed the BH dynamics under the semiclassical approximation. Here we introduce the effect of memory burden that results in a longer PBH lifetime. In order to go beyond the semiclassical picture and invoke the effect of memory burden, we first rewrite Eq.~\eqref{eq:mbh} as
\begin{align}\label{eq:mbh1}
& \mbh(t)=\Min[1-\Gamma_{\rm BH}^0(t-t_{\rm in})]^{1/3}\,,   
\end{align}
where  
\begin{align}\label{eq:Gamma0}
& \Gamma_{\rm BH}^0=3\epsilon\,M_P^4/\Min^3\,, 
\end{align}
is the decay width associated with the PBH evaporation. We assume the semiclassical regime remains valid till
\begin{align}
\mbh=q\,\Min\,.    
\end{align}
In Refs.~\cite{Alexandre:2024nuo,Thoss:2024hsr}, it has been 
proposed that the quantum
effects start to become important when $\mbh=0.5\,\Min$ or $q=0.5$. We consider $t_q$ to be the time scale at which the semiclassical phase ends, beyond which the self-similar regime no longer remains valid. Then, from Eq.~\eqref{eq:mbh1} we obtain 
\begin{align}    
t_q= \f{1-q^3}{\Gamma_{\rm BH}^0}\,.
\end{align}

Once the mass of the PBH reaches $q\,\Min$, the quantum memory effect starts dominating. From this stage onward the BH mass evolution gets modified as
\begin{align}\label{eq:dmdt2}
\frac{d\mbh}{dt}=-\frac{\epsilon}{\left[\mathcal{S}(\mbh)\right]^k}\f{M_P^4}{\mbh^2}\,,
\end{align}
where $S(\mbh)$ is given by Eq.~\eqref{eq:SBH}. Once again, for $\mbh\gg Q/G$, we obtain
\begin{align}
\mbh=q\,\Min\l[1-\Gamma_{\rm BH}^q\,(t-t_q)\r]^{\frac{1}{3+2k}}\,,
\label{eq:mbh2}
\end{align}
with
\begin{align}\label{eq:Gammaq}
\Gamma_{\rm BH}^q= {2^k\,(3+2k)\,\epsilon}\,M_P
\l(\f{M_P}{q\Min}\r)^{3+2k}\,,   
\end{align}
as the modified decay rate and $\tev^q=1/\Gamma^q_{\rm BH}$ as the modified BH lifetime. Clearly, for a given $q$ and $\Min$, a larger $k$ results in a longer BH lifetime. We can now find the modified evolution of BH electromagnetic charge due to the onset of memory burden effect from Eq.~\eqref{eq:dQdt} as,
\begin{align}\label{eq:dQdt2}
& \frac{dQ}{dt}\simeq-Q\,\sum \frac{g_i\,q_i^2}{24\pi}\,\frac{M_P^2}{q\,\Min}\implies Q(t)\simeq \Qin\exp\left[-\sum \frac{g_i\,q_i^2}{24\pi}\,\frac{M_P^2\,t}{q\,\Min}\right]
\,,    
\end{align}
assuming $\mbh\gg Q/\sqrt G$ and $t\gg \tin$. Clearly, decay of the BH charge shall be delayed for $q>0$, however the charge still decays away exponentially as we have seen before [cf. Eq.~\eqref{eq:Qem}]. Once again, same as Eq.~\eqref{eq:betac}, during the memory-burden epoch there exist a critical $\beta$-value~\cite{Masina:2020xhk,Haque:2024eyh} 
\begin{align}\label{eq:betac-mem}
&\beta_c^q=\frac{1}{q}\,\sqrt{\frac{\tin}{\tev^q}}\,,
\end{align}
above which the PBH energy density dominates over the radiation energy density, and a PBH dominated epoch starts until the evaporation completes. Because of the memory burden effect, the BH yield is modified for $\beta>\beta_c^q$, since now the BH evaporation time follows from Eq.~\eqref{eq:Gammaq}, with corresponding evaporation temperature
\begin{align}\label{eq:Tev-MB}
& \Tev^q\Big|_{\beta > \beta_c^q} = M_P\,\l(\f{40}{\pi^2\,\gs(\Tev^q)}\r)^{1/4}\,\l[2^k\,(3+2k)\,\epsilon\l(\f{\Mpl}{q\,\Min}\r)^{3+2k}\r]^{1/2}\,, 
\end{align}
that results in a BH yield 
\begin{align}\label{eq:YBH-mem}
& Y_{\rm BH}(\overline{T}_{\rm ev}^q)=\left(\frac{405}{32}\right)^{1/4}\,\sqrt{\frac{3\epsilon}{\pi}}\,q\,\left(\frac{\gs(\overline{T}_{\rm ev}^q)^3}{\gss(\overline{T}_{\rm ev}^q)^4}\right)^{1/4}\,\left(\frac{M_P}{q\,\Min}\right)^\frac{2k+5}{2}\,\sqrt{2^k\,\left(1+\frac{2k}{3}\right)}\,.    
\end{align}
In deriving Eq.~\eqref{eq:Tev-MB} we have used the fact that in a PBH-dominated universe, $H(t_{\rm ev}^q)=2/\left(3\,t_{\rm ev}^q\right)$. In the opposite limit, i.e., for $\beta<\beta_c^q$, the evaporation temperature is given by
\begin{align}\label{eq:Tev-MB2}
& \Tev^q\Big|_{\beta < \beta_c^q} = M_P\,\l(\f{45}{\pi^2\,\gs(\Tev^q)}\r)^{1/4}\,\l[2^k\,(3+2k)\,\epsilon\l(\f{\Mpl}{q\,\Min}\r)^{3+2k}\r]^{1/2}\,, 
\end{align}
while the yield simply follows Eq.~\eqref{eq:YBH1}. For Eq.~\eqref{eq:Tev-MB2}, we have utilized $H(\tev^q)=1/(2\,\tev^q)$, as the PBH evaporates during radiation domination. Note that, for $k=0,\,q=1$, as expected, Eq.~\eqref{eq:YBH-mem} reproduces Eq.~\eqref{eq:YBH2}.

Now, any modification to Hawking evaporation would also significantly modify the GW associated with its density fluctuation. Following Refs.~\cite{Balaji:2024hpu,Barman:2024iht}, we find an upper bound on initial PBH mass fraction as,
\begin{align}\label{eq:betaGW2}
\beta < 1.2\times 10^{-4}\,q^{-3/4}\,\mathcal{K}(k)\,\exp\left[\frac{3}{8}\,\left(8-28k-\frac{7}{3+2k}\right)\right]
\,\left(q\,\frac{\Min}{1\,\text{g}}\right)^{-\frac{(k+1)\,(14k+17)}{16k+24}}\,,
\end{align}
where $\mathcal{K}(k)=\left(1+\frac{2 k}{3}\right)^{\frac{1}{4 k+6}+\frac{7}{16}} (2k+2)^{-\frac{3}{16 k+24}}$. For $k=0,\,q=1$, we again obtain Eq.~\eqref{eq:betaGW} from Eq.~\eqref{eq:betaGW2}. Due to longer evaporation time because of the memory burden effect, the initial fraction of PBHs must be smaller than in the semiclassical case.
The WDM constraint also gets modified. Following Eq.~\eqref{eq:EDM} and using $\Tbh^q= M_P^2/(q\,\Min)$, we obtain,
\begin{align}
\mdm & \gtrsim 6.8\times 10^{-6}\,\text{GeV} \left(\frac{q\,\Min}{M_P}\right)^\frac{2k+1}{2}\,\frac{\delta}{1.3}\,\frac{1}{\sqrt{2^k\,\epsilon\,(2\,k+3)}}\,\left[\gs(\Tev^q)^\frac{3}{4}\,\frac{\gss(T_{\rm eq})}{\gss(T_{\rm ev}^q)}\right]^\frac{1}{3}\,,    
\end{align}
for $\beta>\beta_c^q$. For a conservative estimation, we chose $\delta\simeq 1.3$ following Refs.~\cite{Lennon:2017tqq,Masina:2020xhk}. Also, we assume there is no entropy injection  from the epoch of evaporation till MRE. Here we only derive an approximate analytical estimation, we will numerically obtain this bound from {\tt FRISBHEE}, that provides more accurate result. 
\section{Results \& discussions}
\label{sec:result}
In the current set-up, after the charged PBHs are formed, they immediately start evaporating via Hawking radiation producing the radiation bath. The total energy density of the Universe is therefore dominated by two components: PBH energy density $\rho_{\rm BH}$ and the radiation energy density $\rho_R$. The time evolution of these energy densities is governed by the Friedmann equations 
\begin{align}
&\frac{d\rho_{\rm BH}}{dt}+3H\rho_{\rm BH}=-\frac{\rho_{\rm BH}}{\mbh}\,\frac{d\mbh}{dt}\,,
\\&
\frac{d\rho_R}{dt}+4H\rho_R=\frac{\rho_{\rm BH}}{\mbh}\,\frac{d\mbh}{dt}\,,
\\&
H=\sqrt{\frac{\rho_{\rm BH}+\rho_R}{3\,M_P^2}}\,,    
\end{align}
along with Eq.~\eqref{eq:dmdt2} and Eq.~\eqref{eq:dQdt} to account for the PBH mass and charge evolution, respectively. To properly track the dynamics of charged PBHs along with memory burden effects, we have implemented our scenario in the publicly available code {\tt FRISBHEE}~\cite{Cheek:2021odj,Cheek:2021cfe,Cheek:2022dbx,Cheek:2022mmy}. In this analysis we consider a monochromatic mass function for the PBHs, and do not include merging or accretion effects \footnote{In Ref.~\cite{Zantedeschi:2024ram} effect of memory burden on PBH merging has been discussed.}.
\begin{figure}[htb!]
\centering
\includegraphics[scale=.8]{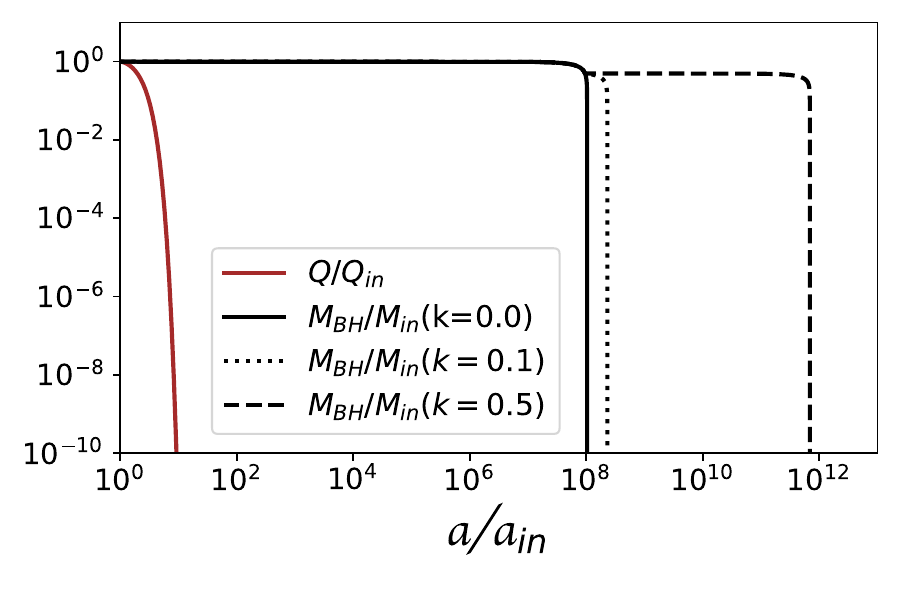}
\caption{Evolution of BH mass and charge with and without the effect of memory burden, with the scale factor for $\Min=10^2$ g, $\beta=10^{-5}$, $q=0.5$, $Q_{\rm in}=0.999$ (in the units of $\sqrt{G}\,\Min$).}
\label{fig:evol}
\end{figure}
\begin{figure}[htb!]
\centering
\includegraphics[scale=.75]{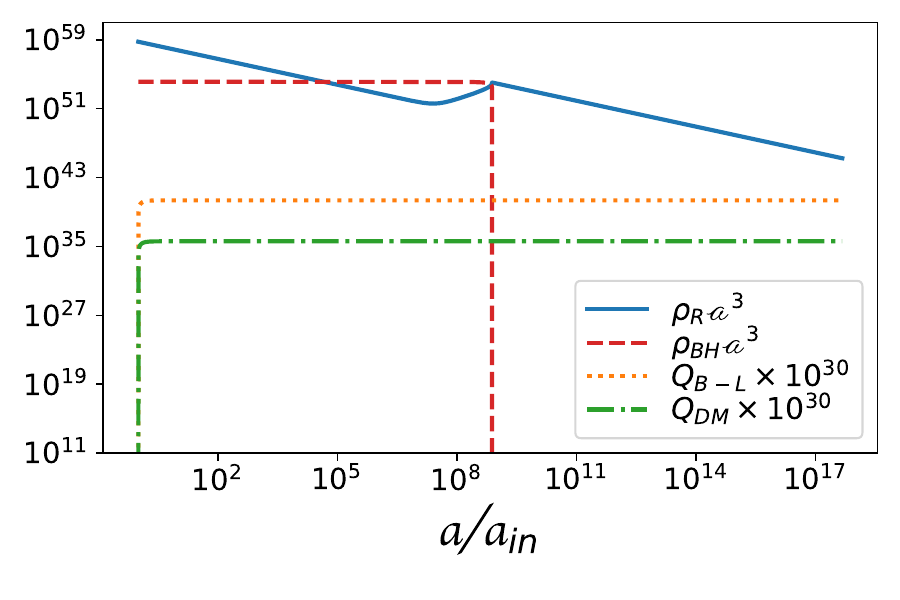}
\caption{Evolution of PBH (red dashed) and radiation (blue solid) energy densities along with $Q_{B-L}$ (orange dotted) and $Q_{\rm DM}$ (green dot-dashed) for a PBH of mass $\Min\simeq 450$ g, with $\beta=10^{-5}$, $\mdm=10^5$ GeV, $\lambda=10^{30}$ and $\Ldm=2.3\times10^{26}$. The asymptotic charges provide observed baryon asymmetry as well as right DM abundance, considering a scalar DM.}
\label{fig:evol_B-L}
\end{figure}
\begin{figure}[htb!]
\centering
\includegraphics[scale=0.75]{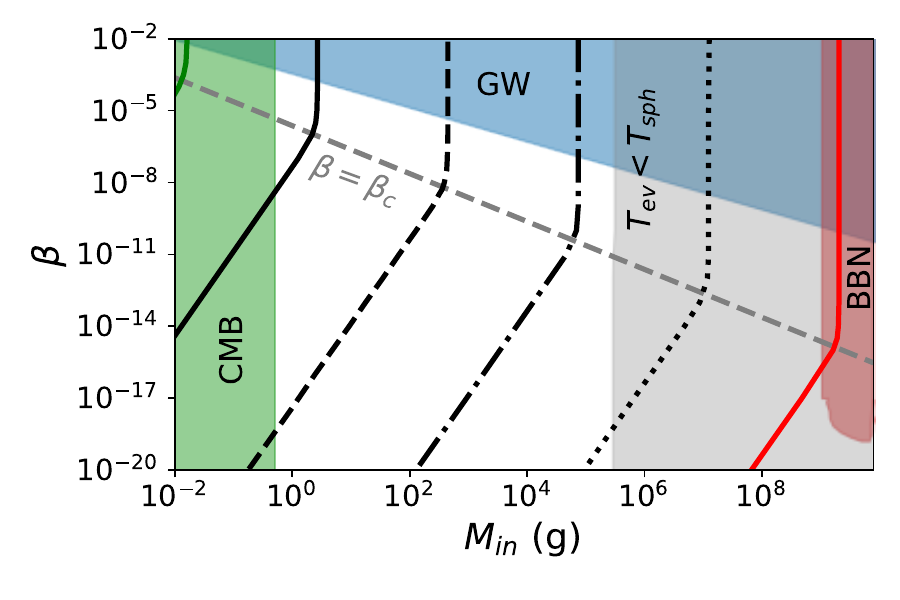}\\
\includegraphics[scale=0.75]{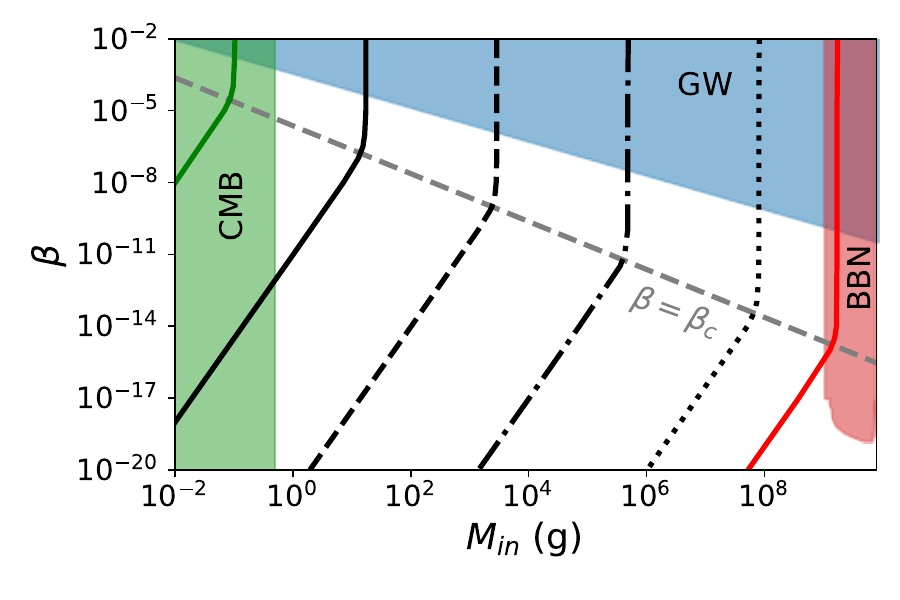}
\caption{Top: Contours providing the observed baryon asymmetry for different choices of $\lambda=\{10^{10},\,10^{20},\,10^{30},\,10^{40},\,10^{50},\,10^{60}\}$ from left to right. The gray shaded region corresponds to $\Tev<160$ GeV (see text for details). Bottom: Same as top, with $\Ldm=\{10^{10},\,10^{20},\,10^{30},\,10^{40},\,10^{50},\,10^{56}\}$ (from left to right), but now the contours correspond to right DM abundance, for a scalar DM of mass $10^5$ g. In both plots the Blue shaded region is disallowed from excessive GW energy density [cf. Eq.~\eqref{eq:betaGW}]. The green shaded region is forbidden from the scale of inflation and the red shaded region is disallowed by BBN bound on PBH lifetime. We consider the semiclassical picture to be always valid.}
\label{fig:contour_B-L}
\end{figure}
\begin{figure}[htb!]
 \centering
 \includegraphics[scale=0.75]{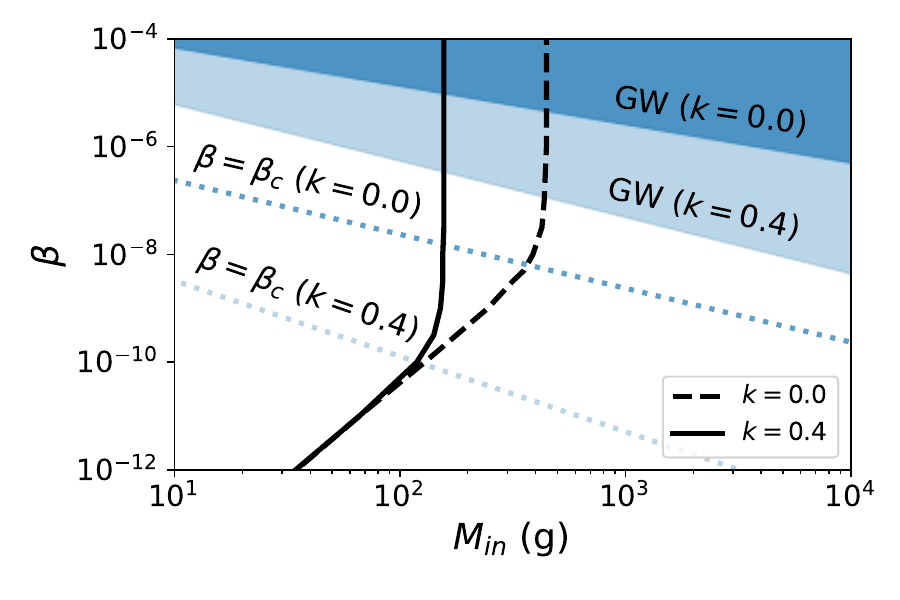}\\ \includegraphics[scale=0.75]{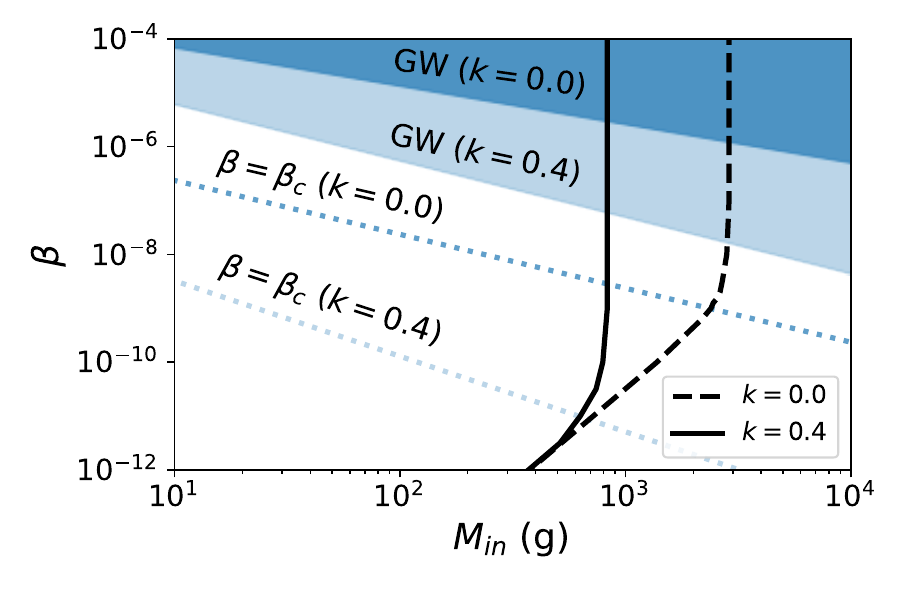}
 \caption{Top: Contours satisfying the observed baryon asymmetry for different choices of $k$ to include the effect of memory burden, for  benchmark values $\lambda=10^{30}$. The dotted straight lines correspond to $\beta=\beta_c$. Bottom: Same as top but the contours correspond to right DM abundance, considering a scalar DM, with $\Ldm=10^{30}$. In both case we consider $q=0.5$. The shaded region is forbidden due to overproduction of GW.}
 \label{fig:contour_B-L_MB}
 \end{figure}
\begin{figure}[htb!]
\centering \includegraphics[scale=0.8]{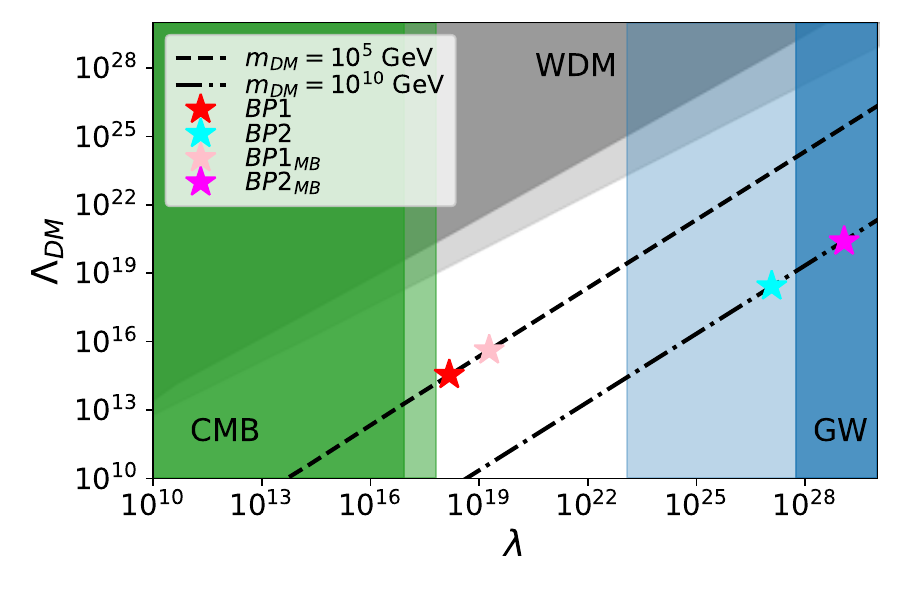}
\caption{The correlation between $\lambda$ and $\Ldm$ in satisfying the observed baryon asymmetry and DM abundance is shown, for a fixed $\beta=10^{-5}$ with $\mdm=\{10^5,\,10^{10}\}$ GeV. Both the semiclassical scenario and the memory burdened scenario follow the same two lines. The darker shaded regions are excluded from the CMB constraint on the scale of inflation (in green), overproduction of GW due to PBH density fluctuation (in blue) and warm DM constraints (in gray) for the semiclassical scenario. The lighter shaded regions are excluded due to memory burden effect for  $q=0.5$ and $k=0.4$. The $\star$ marked points show the benchmarks tabulated in Tab.~\ref{tab:BP}.}
\label{fig:lambdas}
\end{figure}

In Fig.~\ref{fig:evol} we show the evolution of BH mass and electromagnetic charge with time (scale-factor). As expected, for a given $\Min$, inclusion of memory effects $(k\neq 0)$ improve the PBH lifetime by substantial amount, compared to the semiclassical scenario $(k=0)$. However, the same does not hold for $Q$, as it decreases exponentially with time following Eq.~\eqref{eq:dQdt2}. Consequently, before the onset of the memory effect, the charge decays away resulting in no substantial impact on the BH dynamics. Here we have chosen $\beta=10^{-5}$, that ensures BH domination for $\Min=10^2$ g, following Eq.~\eqref{eq:betac}. 

The evolution of $B-L$ and DM charges along with BH and radiation energy densities, as a function of the scale factor is shown in Fig.~\ref{fig:evol_B-L}. For $\beta>\beta_c$, we encounter an intermediate BH domination epoch, as one can see from the red dashed curve. Once the PBH evaporation is completed, the Universe becomes radiation dominated while both $Q_{B-L}$ and $Q_{\rm DM}$ reach constant asymptotic values that correspond to the observed baryon asymmetry, as well as DM abundance. Note that, since post-evaporation, there is no further entropy injection in the bath, the asymmetry remains unchanged. Here we consider the semiclassical treatment to be always valid.    

The parameter space corresponding to observed baryon asymmetry is shown in the top panel of Fig.~\ref{fig:contour_B-L}, assuming the BH to be a semiclassical object. For a given $\lambda$, the curve shows two different slopes for $\beta<\beta_c$ and $\beta>\beta_c$, following Eq.~\eqref{eq:YB0}. For a given $\beta$, heavier PBH requires larger $\lambda$ since in that case longer PBH domination results in dilution of the final asymmetry and hence one needs to go to larger $\lambda$ to obtain the correct asymmetry as $|Q_{B-L}|\propto\lambda$. The allowed mass window for PBH constrains $\lambda$-values to remain within $10^{20}\lesssim\lambda\lesssim 10^{60}$. However, the requirement $\Tev>T_{\rm EW}$ puts a strong bound on the PBH mass, and therefore on $\lambda$, resulting in $10^{20}\lesssim\lambda\lesssim 10^{40}$. We see, our analytical estimation in Eq.~\eqref{eq:lam-YB} matches well with our numerical results. In the bottom panel we illustrate the allowed parameter space that satisfies the DM relic abundance. Once again, we find a the free parameter $\Ldm$ is bounded from above, as well as from below depending on the valid PBH mass window. Finally, for a given PBH mass, there exist upper bound on $\beta$ obtained from Eq.~\eqref{eq:betaGW} in order not to overclose the Universe with excessive GW energy density originating from PBH density perturbation, as indicated by the blue shaded region. For the initial PBH mass considered here, the entire parameter space satisfies $\Tev>T_{\rm EW}$, both for the semiclassical and for the memory-burdened scenario.

We introduce the effect of memory burden, considering $q=0.5$ in Fig.~\ref{fig:contour_B-L_MB}. This implies, the semiclassical approximation breaks down when the PBH mass reaches half of its initial value. We show contours satisfying the observed baryon asymmetry for different choices of $k$ in the top panel, while in the bottom panel different contours produce the total DM abundance. Following Eq.~\eqref{eq:YBH-mem}, we see, for a given $k$, the viable parameter space is affected for $\beta>\beta_c$. Also, a larger $k$ shifts the parameter space to lighter PBHs. This can be understood by inspecting Eq.~\eqref{eq:YBH-mem}, which results in $Y_{\rm BH}\propto 2^k/\Min^{5/2}$ for $2k\ll 3$. As a consequence, the final asymmetry $Y_B^0\propto Y_{\rm BH}$ increases with increase in $k$, that can be compensated by a lighter PBH. In other words, since memory burden enhances the lifetime of PBH, hence it is always preferable to have lighter PBHs to avoid overproduction of the baryon asymmetry.
\begin{table}[htb!]
\centering
\begin{tabular}{|c|c|c|c|c|c|c|c|c|c|c|}\hline
Benchmarks & $\Min$ (g) & $\beta$ & $\mdm$ (GeV) & $k$ & $q$ & $\lambda$ & $\Ldm$ \\
\hline\hline
BP1 & 1 & $10^{-5}$ & $10^{5}$ & 0 & 1 & $1.5\times10^{18}$ & $3.5\times10^{14}$ \\
\hline
BP2 & 100 & $10^{-5}$ & $10^{10}$ & 0 & 1 & $1.2\times10^{27}$ & $2.7\times10^{18}$ \\
\hline\hline
BP1$_{\rm MB}$ & 1 & $10^{-5}$ & $10^{5}$ &0.4&0.5&$1.9\times10^{19}$&$4.2\times10^{15}$ \\
\hline
BP2$_{\rm MB}$ & 100 & $10^{-5}$ & $10^{10}$ & 0.4 &0.5&$1.2\times10^{29}$&$2.5\times10^{20}$\\
\hline
\end{tabular}
\caption{Benchmark points satisfying both observed baryon asymmetry and right DM abundance, following the semiclassical approximation (first two rows) as well as memory burden effect (last two rows).}
\label{tab:BP}
\end{table}

The summary of the allowed parameter space is illustrated in Fig.~\ref{fig:lambdas}, where we show the region of the parameter space in the coupling plane that satisfies the observed baryon asymmetry as well as the DM relic abundance, considering a real scalar DM. We therefore consider the actions in Eq.~\eqref{eq:action1} and \eqref{eq:action-dm} are present simultaneously such that asymmetry is generated in both the visible and in the dark sector from the evaporation of a single BH. In obtaining the net parameter space, we have scanned over the initial PBH mass $\Min\in\left[10^{-2}-10^7\right]$ g, for a fixed $\beta=10^{-5}$. From Eq.~\eqref{eq:lam-YB} and \eqref{eq:lam-DM} it is clear that, one needs $\Ldm>\lambda$ to satisfy both $Y_B^0$ and $\Omega_{\rm DM}\,h^2$. Also, the couplings corresponding to right asymmetries have the same dependence on $\Min$. This is exactly reflected here, for a fixed DM mass. Since a heavier DM requires feebler $\lambda$ to satisfy the observed abundance [cf. Eq.~\eqref{eq:lam-DM}], hence, we see, the contour corresponding to $\mdm=10^{10}$ GeV (black dot-dashed contour) shifts to the lower $\lambda$ values, compared to $\mdm=10^{5}$ GeV (black dashed contour). As smaller couplings correspond to lighter PBHs [cf. Fig.~\ref{fig:contour_B-L}], hence they are constrained from the CMB bound, whereas higher couplings (corresponding to heavier PBHs) are tightly constrained from the bound on $\beta$ due to GW [cf. Eq.~\eqref{eq:betaGW2}]. It is also worth noting that the CMB-disallowed region of the parameter space (green shaded) satisfies $\beta<\beta_c$ since it typically corresponds to $\Min\lesssim 0.7$ g, while the part of the parameter space corresponding to heavier $\Min$, in tension with GW energy density (blue shaded), satisfies $\beta>\beta_c$. We therefore find, the allowed range of coupling to be $10^{17}\lesssim\lambda\lesssim 10^{27}$, while $10^{8}\lesssim\Ldm\lesssim 10^{19}$ for $\mdm=10^{10}$ GeV and $10^{13}\lesssim\Ldm\lesssim 10^{24}$ for $\mdm=10^{5}$ GeV, considering the semiclassical approximation to be valid. In Tab.~\ref{tab:BP} we tabulate a few benchmark points that correspond to right baryon asymmetry and DM abundance. 

Inclusion of memory burden shifts the entire parameter space towards larger couplings along the same contours (this is evident from Fig.~\ref{fig:lambdas}), hence we do not explicitly show them here. However, the CMB and GW bounds now become more stringent, as shown by the lighter shaded regions. Thus, memory-burden effect makes the viable parameter space narrower. Finally, we project the WDM limit obtained numerically from {\tt FRISBHEE}, following the methodology in Ref.~\cite{Baur:2017stq} (as described in Sec.~\ref{sec:DM}). This rules out the parameter space corresponding to DM mass of $\mdm\simeq\left[10^{-2}-3\times 10^2\right]$ GeV, as shown via the gray shaded region. 
\section{Conclusions}
\label{sec:concl}
The dynamical generation of the matter-antimatter asymmetry and the identification of a viable particle candidate for dark matter (DM) remain enduring challenges in modern physics.  In this work, we explore a purely gravitational origin for both phenomena. Given that gravity is an inescapable interaction, this approach provides a fundamental and compelling pathway. Our baryogenesis mechanism hinges on a quantum-gravity-inspired coupling between spacetime and the $B-L$ current. This coupling introduces a chemical potential that discriminates between baryons and antibaryons, thereby driving the observed baryon asymmetry. In presence of such a chemical potential, the Hawking evaporation from a primordial black hole (PBH) becomes asymmetric.

We consider PBHs formed in a radiation-dominated epoch that evaporate before the onset of Big Bang Nucleosynthesis (BBN), implying they are relatively light. Our findings indicate that the observed baryon asymmetry can be generated for such light PBHs with a coupling strength of $\gtrsim\mathcal{O}(10^{20})$ [cf. Fig.~\ref{fig:contour_B-L} upper panel]. Similarly, we propose that an analogous coupling between spacetime and a current in the dark sector could account for the observed DM abundance via asymmetric DM generation [cf. Fig.~\ref{fig:contour_B-L} lower panel]. Beyond the semiclassical Hawking radiation framework, we incorporate the memory-burden effect to explore the parameter space that simultaneously satisfies the observed baryon asymmetry and asymmetric DM abundance [cf. Fig.~\ref{fig:contour_B-L_MB}]. Our analysis outlines the viable parameter space consistent with constraints from CMB on the inflationary scale, BBN limits on PBH lifetimes, and the production of primordial gravitational waves arising from PBH density fluctuations [cf. Fig.~\ref{fig:lambdas}]. In conclusion, this study presents a purely gravitational framework for the generation of asymmetries in both the visible and dark sectors, offering a unified perspective.

\section*{Acknowledgement}
BB would like to acknowledge email communications with 
Nolan Smyth.  OZ has been partially supported by Sostenibilidad-UdeA, the UdeA/CODI Grants 2022-52380 and 2023-59130, and the Ministerio de Ciencias Grant CD 82315 CT ICETEX 2021-1080.
\bibliography{Bibliography}
\bibliographystyle{JHEP}
\end{document}